# Necessary and sufficient condition for a disorder-broadened transition to be identified as 1<sup>st</sup> order

P Chaddah*

RR Centre for Advanced Technology, Indore 452013, India.

*Disorder-broadened 1$^{st}$ order transitions may not carry the experimental signature of a latent heat. We discuss what experimental observation can provide a necessary and sufficient condition for characterizing this phase transition as 1$^{st}$ order.*

A phase transition is observed when a control variable is varied monotonically through a critical value. The variable used most commonly is temperature T; pressure P, and magnetic field H, are also often used. A phase transition is said to have occurred when some measurable quantity (measurable quantities are derivatives of the free energy) changes drastically. If the quantity that changes discontinuously is a 1$^{st}$ derivative of the free energy (like density, or magnetization, or entropy) then it is called a 1$^{st}$ order transition. If these quantities are continuous, but the measurable quantity that changes discontinuously is a higher derivative of the free energy, then it is classified as a 2$^{nd}$ order transition. In the text-book definition of a 1$^{st}$ order phase transition, the entropy S changes discontinuously and a *Latent Heat* must be associated with the transition. Further the 1$^{st}$ derivative of free energy associated with the second control variable viz. density or magnetization also changes discontinuously. In a 2$^{nd}$ order phase transition there is no latent heat; entropy, and density or magnetization are continuous, and it is some higher derivative of free energy that is discontinuous. Observation of a latent heat is considered as a necessary and sufficient condition for characterizing a phase transition as 1$^{st}$ order.

As I start discussing the case of disorder-broadened 1$^{st}$ order transitions [1] we will find that these definitions may no longer be tenable. Imry and Wortis [2] had discussed the influence of increasing disorder on a 1$^{st}$ order transition, and this is depicted in the schematic figures 1 to 3. Does a 1$^{st}$ order transition that is broadened by disorder always show a latent heat? Does the 1$^{st}$ derivative of free energy always show one discontinuity at the transition (see figure 1 and 2), and does observation of a latent heat remain a necessary and sufficient condition for characterizing a phase transition as 1$^{st}$ order? Or can a disorder-broadened 1$^{st}$ order transition show a series of discontinuities, as depicted schematically in figure 4? This corresponds to different regions of the sample, each of the length scale of the correlation length, having different $T_C$ because of having different distribution of quenched impurities [1]. The number of steps will increase, and the step-size will decrease, as the correlation length reduces or the number of such regions increases; an experiment can no longer identify a latent heat at a particular value of the control variable. Since the correlation length is dictated by the range of the interaction driving the phase transition, this can be much shorter than the length scale over which other properties (like say conduction electron density measured in a photoemission experiment) are decided. The sample can appear homogeneous when say scanning PES is used, and can show a $2^{nd}$ order paramagnetic to ferromagnetic transition with an infinite correlation length, but show a broad $1^{st}$ order ferromagnetic to anti-ferromagnetic transition occurring over a range of temperatures because this transition is driven by a very short-range interaction.

The scenario depicted in figure 4 was observed experimentally by Soibel et al [3] in a vortex-lattice melting transition, and by Roy et al [4] in a ferromagnetic to anti-ferromagnetic transition. Both these studies showed that the transition occurs over small regions at different values of the control variable (magnetic field H in both these studies) as it scans through the phase transition line. A large number of $1^{st}$ order transitions have been studied and understood following the schematic of figure 4, as has been reviewed in recent works [1, 5].

In the absence of the experimental signature of a latent heat, we need an experimental signature defining the $1^{st}$ order nature of the transition. The existence of hysteresis has been used as a distinguishing characteristic of a $1^{st}$ order transition, but it will be observed only if the energy associated with fluctuations is smaller than a critical value (corresponding to the height of the free energy barrier separating the two local minima). Thus observation of hysteresis is a sufficient condition, but not a necessary condition, for characterizing a transition as $1^{st}$ order. This needs to be emphasized because there are many experimental reports where the thermal hysteresis in a disorder-broadened $1^{st}$ order transition reduces as the second control variable (usually magnetic field H) is used to shift the transition temperature higher [6]. The extrapolated value of H, where hysteresis is no longer observable, is termed as the critical end-point where the phase transition becomes $2^{nd}$ order. While this reduction in hysteresis with rising $T_C$ has been argued as generic [1,7], and its observation been reported as consistent with this phenomenological prediction [8,9], this prediction has been ignored [6]. As argued above, hysteresis cannot be observed in a $2^{nd}$ order transition. This makes its observation a sufficient condition for characterizing a phase transition $1^{st}$ order. However, hysteresis will not be observed across a $1^{st}$ order transition if the energy associated with fluctuations is larger than the height of the free energy barrier separating the two local minima. Accordingly, observation of hysteresis is not a necessary condition for a $1^{st}$ order transition. Not observing hysteresis is not enough to characterize a phase transition as $2^{nd}$ order.

Here I will distinguish 1$^{st}$ and 2$^{nd}$ order phase transitions through the process by which the phase transition occurs. We shall present characteristics of a 1$^{st}$ order transition that must be valid for the case of no disorder, with the proviso that these characteristics remain even for disorder-broadened 1$^{st}$ order transitions of the type depicted in figure 4.

A system in equilibrium at a given T and P (or a given T and H) is in a minimum of the Gibbs free energy defined as G = U+**P**V-**T**S (or U-**H**M-**T**S). It follows that if a phase transition occurs on a ($T_C$, $P_C$) line, or on a ($T_C$, $H_C$) line, then the two phases A and B that exist on either side of the line have the **same** value of G on this line i.e. $G_A=G_B$ on the phase transition line. It also follows that the inequality between $G_A$ and $G_B$ changes sign as this line is crossed, since the equilibrium phase has the lowest G. We now discuss how phase transitions, both 1$^{st}$ order and 2$^{nd}$ order, proceed as we attempt to cross this line.

For a 2$^{nd}$ order phase transition, entropy and density change continuously across the ($T_C$, $P_C$) line [or entropy and magnetization change continuously across the ($T_C$, $H_C$) line]. Since the Gibbs free energy is continuous across all phase transitions, it follows that for a 2$^{nd}$ order phase transition the internal energy U also changes continuously. Thus in addition to entropy and density, the internal energy U for the two phases is also equal on the ($T_C$, $P_C$) line [or on the ($T_C$, $H_C$) line] in a 2$^{nd}$ order phase transition. The two phases cannot be distinguished on the phase transition line. The internal energies of the two phases also being equal, an infinitesimal energy change (caused by an attempt to change any of the control variables T, P, or H) will cause transformation of one phase to the other across the whole volume of the sample. This is stated as 'the correlation length for a 2$^{nd}$ order transition is infinite'. The entire sample is either in phase A or in phase B, and there is no question of phase A and phase B coexisting. Phase coexistence cannot be observed in a 2$^{nd}$ order transition.

We contrast this with the case of a 1$^{st}$ order phase transition. Here entropy and density (or magnetization) change discontinuously across the phase transition line, and are different in the two phases on the ($T_C$, $P_C$) line [or on the ($T_C$, $H_C$) line]. The internal energy U is also different in the two phases on the ($T_C$, $P_C$) line [or on the ($T_C$, $H_C$) line]. The two phases can clearly be distinguished on the phase transition line. The internal energies of the two phases being different, a small energy change (caused by an attempt to change any of the control variables T, P, or H) cannot cause transformation of one phase to the other across the whole volume of the sample; it can only transform a small region whose volume is dictated by the energy provided. (The correlation length for a 1st order transition is finite.) We cannot transform a region of length scale smaller than the correlation length; this provides a lower bound on the "critical radius for nucleation" [10]. For a pure system with no disorder, phase coexistence occurs on the ($T_C$, $P_C$) line [or on the ($T_C$, $H_C$) line], and the control variable can take a value away from this line only when the phase transformation has been completed. Thus phase coexistence is essential for a 1st order phase transition to proceed. The observation of the two coexisting phases, albeit only on the ($T_C$, $P_C$) line [or on the ($T_C$, $H_C$) line], is an accepted characteristic of a 1st order transition in a system with no disorder. Let us now consider a system with disorder, where the 1st order transition is driven by short-range interactions, corresponding to the schematic of figure 4. Here the two phases will coexist over the range of the control variable over which the steps are observed, with the phase being transformed in different regions of the sample [11, 12]. Observation of phase coexistence, as the transformation proceeds over the broad temperature window over which the transition is seen, is essential. Combining this with the last sentence of the previous paragraph, observation of phase coexistence is a necessary and sufficient condition for characterizing a phase transition as 1st order.

I now wish to make a comment on the role of impurities, in the form of quenched disorder, in the transformation process of the 1st order transition. As the transition proceeds through nucleation and growth, these impurities are generally believed to act as pins that inhibit the growth process. The schematic of figure 4, and the discussions in references [11,12], view the impurities as changing the local transition temperatures. This influences nucleation, rather than influencing growth, in the 1st order transformation process. Various observations on the devitrification of magnetic glasses [1,5] correspond to temperatures well below the thermodynamic $T_C$ where the critical radius for nucleation is very small. When this radius is large then growth takes preference over nucleation for small energy inputs, as in the case of growth of single crystals. In the devitrification studies, nucleation would appear to be the preferred transformation mode. Phase coexistence would, of course, be the characteristic of the 1st order transition for both modes of the transformation process.

*Since retired; Email: chaddah.praveen@gmail.com

**References:**

1. P Chaddah, arXiv:1403.6319 (2014).
2. Y Imry, and M Wortis, **Phys. Rev. B19**, 3580 (1979).
3. A Soibel, E Zeldov, M Rappaport, Y Myasoedov, T Tamegai, S Ooi, M Konczykowski and V B Geshkenbein, **Nature (London) 406**, 283 (2000).
4. S B Roy, G K Perkins, M K Chattopadhyay, A K Nigam, K J S Sokhey, P Chaddah, A D Caplin, and L F Cohen, **Phys Rev Lett 92**, 147203 (2004).
5. S B Roy and P Chaddah, arXiv:1401.0891 (2014).
6. D Mohan Radheep et al, **JMMM** (2014, in press); P Sarkar et al, **Phys Rev Lett 103**, 057205 (2009); **Phys Rev B 79**, 144431 (2009); L Demko et al, **Phys Rev Lett 101**, 037206 (2008).
7. P Chaddah, **Pramana-J.Phys 67**, 113 (2006); arXiv:0602128 (2006).
8. B I Belevetsev et al, **Phys Rev B74**, 054427 (2006).
9. R R Doshi et al**, Physica B 406**, 4031 (2011).
10. E. M. Lifshitz and L. P. Pitaevski, Landau and Lifshitz Course of Theoretical Physics, Vol. 10, Physical Kinetics (Pergamon Press, Oxford, 1981), Chap. 12; P. M. Chaikin and T. C. Lubensky, Principles of Condensed Matter Physics(Cambridge University Press, Cambridge, 1994).
11. P Chaddah, A Banerjee and S B Roy, arXiv:0601095 (2006); K Kumar et al, **Phys Rev B73**, 184435 (2006).
12. F Macia et al, **Phys Rev B76**, 174424 (2007).

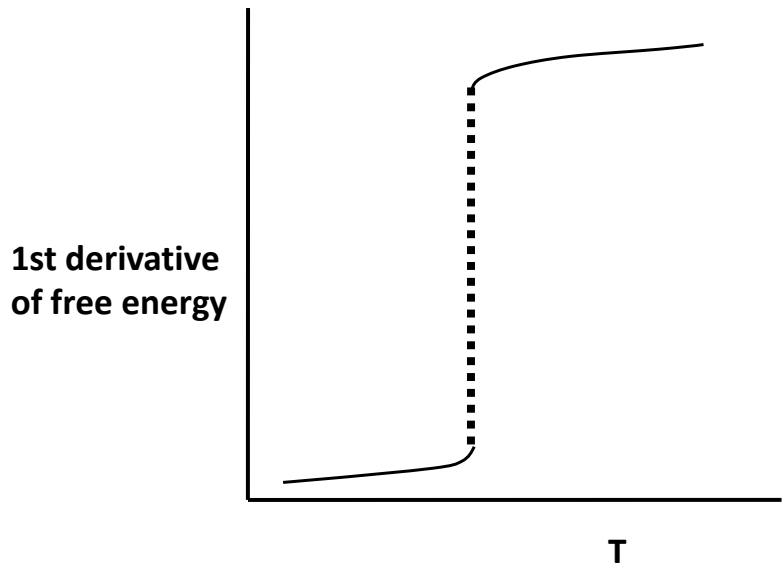

Figure 1: This schematic shows the characteristic sharp discontinuity (at $T_C$) in the 1st derivative of the free energy that is expected in a pure system.

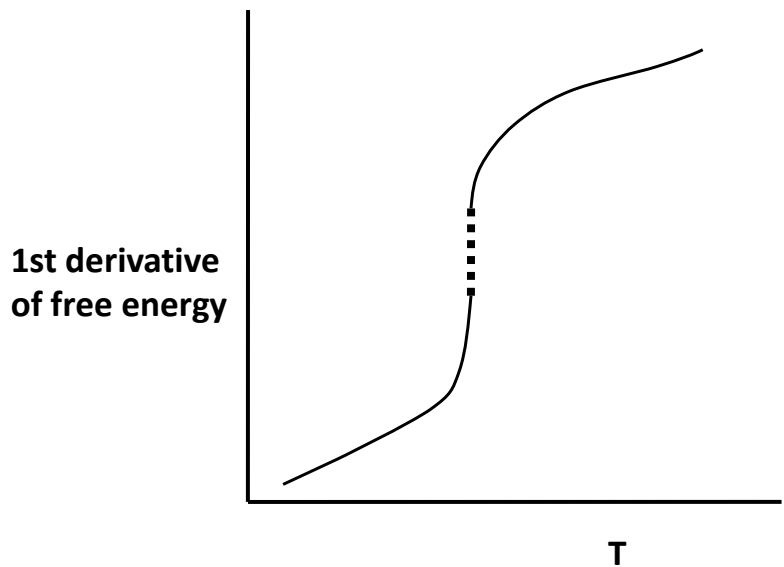

Figure 2: This schematic shows a partial rounding of the 1st order transition due to the presence of disorder, as discussed by Imry and Wortis [2]. The characteristic sharp discontinuity in the 1st derivative of the free energy is markedly reduced from that expected in a pure system, and is seen only at $T_C$.

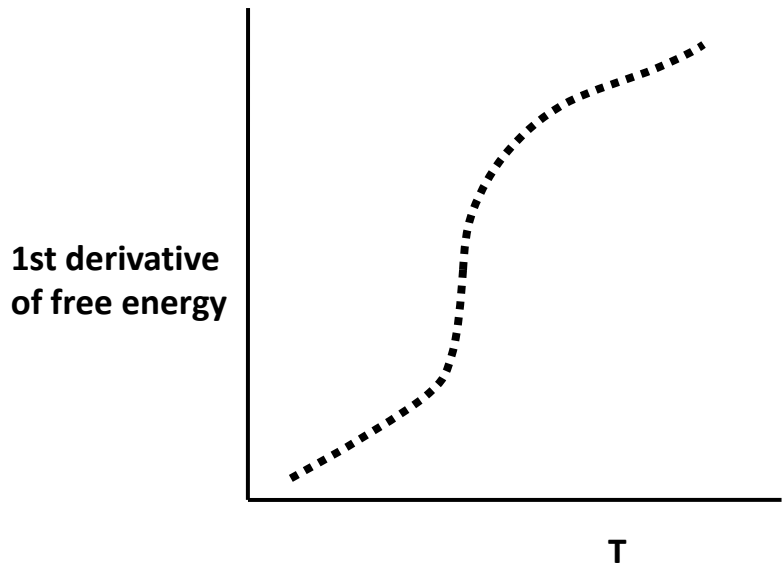

Figure 3: With increasing disorder there is a complete rounding of the transition, and the behavior of the 1st derivative of free energy mimics a 2nd order transition.

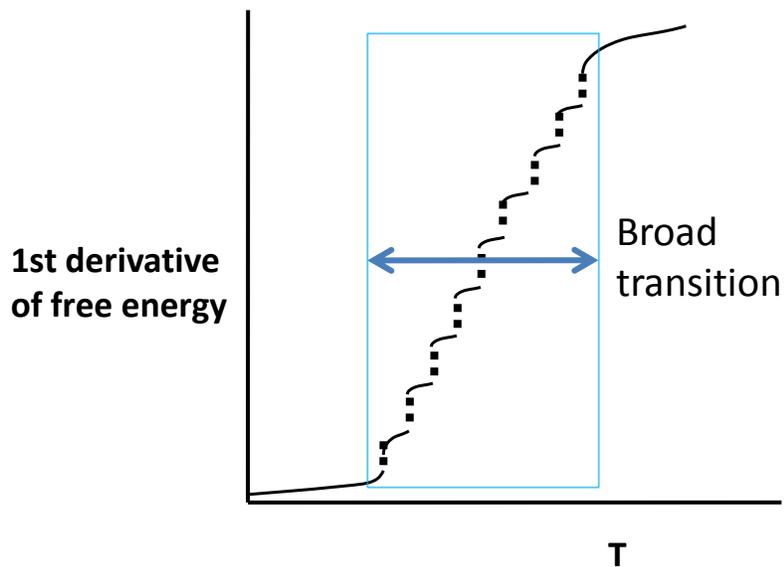

Figure 4: This schematic shows a distribution of the 1st order transition temperatures due to the presence of disorder, as was reported experimentally [3,4], but with magnetic field H as the control variable. The number of steps observed will increase as the correlation length for the transition, or the size of the nuclei of the new phase, reduces. This can be expected when the transition is caused by short range interactions. The characteristic sharp discontinuity in the 1st derivative of the free energy is replaced by a series of small discontinuities, whose magnitude becomes smaller as the size of each independently nucleating region becomes smaller [11,12].